\documentstyle[12pt,aaspp4]{article}

\newcommand{\etal}{{\it et al.\,}}

\received{7/12/96}
\revised{}
\accepted{7/29/97}
\journalid{}{}
\articleid{}{}
\paperid{}
\ccc{}

\begin{document}

\title{X-ray Observations of Distant Optically Selected Clusters}

\author{B. P. Holden}
\authoraddr{Department of Astronomy and Astrophysics, University of
Chicago, 5640 South Ellis Ave. Chicago, Illinois 60637}
\affil{Department of Astronomy and Astrophysics, University of
Chicago, 5640 South Ellis Ave. Chicago, Illinois 60637}
\affil{holden@oddjob.uchicago.edu}

\author{A. K. Romer} 
\authoraddr{Department of Physics, Carnegie Mellon University, 
5000 Forbes Ave. Pittsburgh, Pennsylvania 15213-3890} 
\affil{Department of Physics, Carnegie Mellon University,
5000 Forbes Ave. Pittsburgh, Pennsylvania 15213-3890}
\affil{romer@astro.phys.cmu.edu}

\author{R. C. Nichol}
\authoraddr{Department of Physics, Carnegie Mellon University, 
5000 Forbes Ave. Pittsburgh, Pennsylvania 15213-3890} 
\affil{Department of Physics, Carnegie Mellon University,
5000 Forbes Ave. Pittsburgh, Pennsylvania 15213-3890}
\affil{nichol@astro.phys.cmu.edu}

\author{M. P. Ulmer} \authoraddr{Department of Physics and
Astronomy, Northwestern University, Dearborn Observatory, 2131 Sheridan
Road, Evanston, Illinois 60208-2900}
\affil{Dearborn Observatory, Northwestern University, 2131 Sheridan
Road, Evanston, Illinois 60208-2900} 
\affil{m-ulmer2@nwu.edu}

\begin{abstract}

We have measured fluxes or flux limits for 31 of the 79 cluster
candidates in the Palomar Distant Cluster Survey (PDCS) using archival
ROSAT/PSPC pointed observations.  Our X-ray survey reaches a flux
limit of $\simeq 3 \times 10^{-14}$ erg s$^{-1}$ cm$^{-2}$ (0.4 - 2.0
keV), which corresponds to luminosities of $L_x\simeq 5 \times
10^{43}$ erg s$^{-1}$ (${\rm H_o}$ = 50 km s$^{-1}$ Mpc$^{-1}$, ${\rm
q_o}$ = $\frac{1}{2}$), if we assume the PDCS estimated redshifts.  Of
the 31 cluster candidates, we detect six at a signal-to-noise greater
than three. We estimate that $2.9^{+3.3}_{-1.4}$ (90\% confidence
limits) of these six detections are a result of X-ray emission from
objects unrelated to the PDCS cluster candidates.  The net surface
density of X-ray emitting cluster candidates in our survey,
$1.71^{+0.91}_{-2.19}$ clusters deg$^{-2}$, agrees with that of
other, X-ray selected, surveys. It is possible, given the large error
on our contamination rate, that we have not detected X-ray emission
from any of our observed PDCS cluster candidates.  We find no
statistically significant difference between the X-ray luminosities of
PDCS cluster candidates and those of Abell clusters of similar optical
richness. This suggests that the PDCS contains objects at high
redshift similar to the low redshift clusters in the Abell
catalogs. We show that the PDCS cluster candidates are not bright
X-ray sources, the average luminosity of the six detected candidates
is only $\bar{L_x}=0.9\times10^{44}$ erg s$^{-1}$ (0.4-2.0 keV). This
finding is in agreement with previous X-ray studies of high redshift,
optically selected, rich clusters of galaxies.

\end{abstract}

\section{Introduction}

One of the focal points of modern observational cosmology is the study
of how structure evolves. Clusters of galaxies provide an excellent
probe of such evolution because they represent the largest
gravitationally bound objects and can be identified over a large range
of redshifts. X-ray surveys for clusters of galaxies have produced
catalogs of physically massive objects, but most high redshift X-ray
surveys are limited to small ($\sim 30$) samples (\cite{henry92} 1992;
\cite{castander95} 1995; \cite{collins97} 1997).  By contrast, the
availability of large format CCDs and sophisticated cluster finding
algorithms means that it is now possible to produce optically selected
catalogs, such as the Palomar Distant Cluster Survey (PDCS,
\cite{postman96} 1996), which contain $\sim 100$ distant clusters with
a large range of richnesses. In this paper we seek to compare the
distant clusters found in automated optical surveys with those found
with more traditional methods by studying the X-ray properties of the
PDCS and by comparing the PDCS to lower redshift cluster samples.
  
Previous studies (\cite{henry82} 1982; \cite{bower94} 1994;
\cite{castander94} 1994; \cite{nichol94} 1994; \cite{sok96} ) have
demonstrated that optically rich clusters at high redshift are often
X-ray faint.  Many of the clusters examined have velocity dispersions
consistent with massive, bound, systems (\cite{bower97} 1997). The low
luminosity of these systems has been cited as evidence for X-ray
evolution in the cluster population (\cite{bower94} 1994).  However,
this interpretation may be premature, since the high redshift optical
samples studied to date are extremely small and heterogeneously
selected.

With our survey we are able to improve on previous X-ray studies of
high redshift, optically selected, clusters. Our sample has the
combined advantages of size (31 clusters in all), an objective cluster
catalog (the PDCS) and random selection for the X-ray observations.
We are also able to take advantage of two recent, low redshift
($z<0.2$) X-ray surveys (\cite{bh93} 1993; \cite{burg94} 1994) of
Abell clusters (\cite{abell58} 1958; \cite{aco89} 1989) which are both
statistically complete and contain a large ($\sim 200$) number of
objects.  Our sample should provide a fair representation of the X-ray
properties of the PDCS as a whole and allow us to make direct
comparisons with the \cite{bh93} (1993) and \cite{burg94} (1994)
surveys. Our survey will help us understand what sorts of galaxy
overdensities were selected as clusters by the PDCS and enable us to
test for evolution in the cluster population.
 
In Sec. 2, we review the properties of the Palomar Distant Cluster
Survey.  Section 3 describes our analysis of the fourteen archival
ROSAT/PSPC pointings that contain our X-ray data.  We describe
possible sources of contamination and compare with other published
work in Sec. 4.  We discuss our results in Sec. 5, and make
conclusions in Sec. 6. Throughout this paper we assume ${\rm H_o}$ = 50 km
s$^{-1}$ Mpc$^{-1}$ and ${\rm q_o}$ = $\frac{1}{2}$.

\section{The Palomar Distant Cluster Survey}

The PDCS (\cite{postman96} 1996) was carried out using the 4-Shooter
CCD Camera on the Hale 5m telescope in both the $I_4$ and $V_4$ bands
(\cite{gunn_4s}).  The survey was compiled from scans in five
different regions of the sky and covers a total of 5.1 deg$^{2}$.  The
five fields were selected to have high Galactic latitude, low
reddening and to avoid known, low redshift ($z<0.1$),
clusters. Throughout this paper we will refer to the fields by their
respective hour of right ascension, {\it i.e.} $\alpha$ = 00${\rm
^h}$, 02${\rm ^h}$, 09${\rm ^h}$, 13${\rm ^h}$, \& 16${\rm ^h}$.  The
feature that sets the PDCS apart from previous searches for high
redshift clusters ({\it e.g.}  \cite{gunn86} 1986; \cite{couch91}) is
that it uses an automated matched filter technique to select cluster
candidates from the galaxy catalog.  This objective technique
contrasts with the eye-ball methods used in the past and allows the
authors to generate an accurate selection function based on
simulations.

The PDCS catalog contains 79 entries and for each entry
\cite{postman96} (1996) present several parameters. These include
estimates of radius, redshift and richness. We make use of the PDCS
estimated redshifts in Sec. 3 to calculate luminosities and to help
define metric apertures within which source fluxes are measured. The
estimated redshifts carry a large error, $\delta_z$ = 0.2, and we
discuss in Sec. 5 what impact these errors have on our
conclusions. Also in Sec. 5, we make use of the PDCS matched filter
richness estimate, $\Lambda_{cl}$, to examine the relationship between
optical and X-ray cluster properties.  We note that, for clusters with
redshifts less than $z=0.7$, the matched filter richness can be
equated with the Abell richness (see Sec. 5 and Figure 21 of
\cite{postman96} 1996).

\cite{postman96} (1996) have carried out extensive simulations to
derive an estimate of the completeness and contamination of their
catalog.  They generated artificial galaxy maps designed to reproduce,
as closely as possible, the background galaxy distribution in the five
PDCS fields. The rate of false positive detections in these maps
should represent the rate of contamination in the PDCS catalog. They
plot, in Figure 18 of their paper, the false positive rate as a
function of both measured size and detection significance.
\cite{postman96} (1996) estimate that the contamination level in the
PDCS sample is at the 12\%$-$31\% level. They used a different set of
simulations to investigate the completeness of their survey. They
simulated $\gtrsim 10000$ clusters over a wide range in redshift,
richness and projected radial profiles and conclude that, for clusters
with Abell richnesses $\ge 1$, their catalog is complete over the
range $0.2 \le z \le 0.6$.  Their simulations included clusters that
did not match their matched filter in radial profile, but all models
were spherically symmetric.

In summary, the simulations carried out by \cite{postman96} (1996)
have shown that the PDCS provides a sensitive way to identify optical
cluster candidates.  However, the simulations also highlight the need
for optical spectroscopy and X-ray imaging to confirm the identity of
the objects in the PDCS.  We note that, since up to $31\%$ of the
objects in the PDCS catalog might not be not real clusters, we will
refer to the entries in this catalog as `PDCS cluster candidates'
throughout this paper.

\section{The X-ray Data Analysis}

We have searched the High Energy Astrophysics Science Archive Research
Center (HEASARC) database for those ROSAT/PSPC pointings which have
PDCS cluster candidates in their field of view.  Fourteen such
pointings were found and these are summarized in Table
\ref{pointings}.  We note that, at the time of writing, no ROSAT PSPC
pointing data were available for the PDCS fields at 02${\rm ^h}$ or
16${\rm ^h}$.
  
We reduced all fourteen pointings using the Extended X-ray Analysis
Software (EXAS) of \cite{snowden94}.  EXAS corrects for the energy
dependence of the vignetting function and the particle
backgrounds. EXAS also removes time intervals where the total count
rate was too high ($>5.7$ cts s$^{-1}$) or where an event was
contaminated by a previous pulse. The software produces exposure maps,
count maps and particle background maps in seven different energy
bands (R1 through R7). For our analysis, we combined the maps in bands
R3 through R7 to produce a single count-rate map and error map in the
energy range 0.4 - 2.0 keV for each pointing. We excluded the lowest
bands, R1 and R2, which cover the energy range 0.1 - 0.4 keV, to
minimize the effects of absorption by Galactic hydrogen.

For a PDCS cluster candidate to be included in our sample, we required
that its optical centroid be less than $40'$ from the PSPC pointing
center.  Moreover, the PSPC exposure time at the optical centroid had
to exceed 3000 seconds. Thirty-one PDCS cluster candidates met these
requirements.  For each of candidates, we derived an aperture for the
flux measurement using a cluster model based on a modified isothermal
sphere:
\begin{equation} 
I = \frac{I_o}{[1+(r/r_c)^2]^{3\beta - 1/2}}\,\,\,\,\,, 
\label{I_model}
\end{equation}
where $I$ is the surface brightness at radius $r$.  We used values for
the slope (\( \beta =\frac{2}{3} \)) and core radius (\(r_c = 250\)
kpc) which are typical for rich clusters (\cite{jonesforman92}).  We
converted the above model from physical units to angular units using
the estimated redshifts from the PDCS unless a spectroscopic redshift
was available.  We then convolved the model with an appropriate PSPC
point spread function (PSF), using the PSF model derived empirically
by \cite{nichol94} 1994.  We increased the radius of the aperture
until the integrated flux within the aperture included 70\% of the
total flux of the PSF convolved model.  This radius depends on the
size of the PSF, but is approximately $3.5 \times r_c$. The aperture
radii are listed in column five of Table \ref{rawdata} and range from
1.8\arcmin\ to 3.6\arcmin.
 
Before measuring fluxes, we masked out certain aperture pixels.  We
masked those pixels that were common to more than one cluster aperture
and those pixels that had less than 3000 seconds of exposure. (The
latter tend to fall in the shadow of the PSPC window support
structure.)  We also ran a source detection algorithm on the 14
pointings. This allowed us to mask any pixels that contained flux from
sources with centroids more than 2\arcmin\ (four times the uncertainty
in the PDCS positions) from the PDCS cluster candidates.  (The
detection software provided outlines of the sources as well as
centroid information.) We note that in two cases, PDCS 36 \& 62, a
source was detected within 2\arcmin\ of the PDCS candidate and that
the pixels associated with these two sources were not masked.  The
source detection algorithm we used was that of the Serendipitous
High-redshift Archival Cluster (SHARC) Survey and is more completely
described in \cite{freeman95} and \cite{nichol97} 1997.

In total, 16 of the 31 apertures were masked in some way. In four
cases, more than $\simeq 50\% $ of the aperture was masked. The
fraction of the aperture left available for flux derivation is listed
in column nine of Table \ref{rawdata}.  For those clusters covered by
multiple pointings (all those at $\alpha$ = 09$^{\rm h}$), this column
lists the maximum aperture fraction available from among the
individual observations. The count-rate for each of the 31 PDCS
cluster candidates was measured by summing the flux in the unmasked
aperture pixels. The corresponding background count-rates were
measured in annuli covering 1.5 times the area of, and surrounding,
the unmasked apertures. The background apertures were masked, where
appropriate, following the rules described above.

The PDCS cluster candidates in the collection of pointings at $\alpha$
= 09$^{\rm h}$ required special treatment. These 17 objects were observed
multiple times (see Table \ref{rawdata} column seven) but always at
different positions on the PSPC.  To maximize the effective amount of
exposure time for these clusters, we combined the (background
subtracted) count-rate measurements from each observation via a
weighted average to produce a mean, background subtracted, count-rate
for each source. The weights were the product of the average exposure
time and the fraction of the total aperture used.

With all background subtracted count-rates in hand, we then determined
which of the 31 observations could be classified as detections. For
this we used the (0.4 - 2.0 keV) error map from the EXAS package to
compute the error, in counts per second, in each (masked)
aperture. (The errors for the $\alpha$ = 09${\rm ^h}$ objects were derived
by combining the errors on the individual observations in a weighted
fashion.) We classified any observation as a detection if the ratio of
the background subtracted count-rate to the error was greater than
three \footnote{We note here that this paper uses a different
definition of detection than the SHARC Survey.}  Six
observations met this criterion and they are marked with an asterisk
in Table \ref{rawdata}, column one, and illustrated in Figure
\ref{mosaic}.  Also in Table \ref{rawdata}, we list the background
subtracted count-rate (column eight) and the signal-to-noise ratio of
the detection (column ten).  For the 25 PDCS cluster candidates that
were not detected, we list an upper limit to their background
subtracted count-rate given by three times the measured error.

We converted our measured count-rates (or 3$\sigma$ upper limits) to
energy fluxes by integrating a redshifted 6.0 keV thermal {\it
bremsstrahlung} spectrum over our energy passband of 0.4 - 2.0 keV.
All fluxes were then corrected for absorption using the observed
amount of Galactic neutral hydrogen in the AT\&T Bell Laboratories 21
cm survey (\cite{stark}) and the cross-section values from
\cite{mm83}.  Once we had the energy flux inside the masked aperture
(column two of Table \ref{processedstuff}), we needed to convert to a
total flux for each cluster candidate.  For cluster apertures where no
part of the aperture was masked, the energy flux was simply divided by
0.7 (see previous discussion) to give a total flux (column four of
Table \ref{processedstuff}).  For the apertures that were masked, we
computed the fraction of the flux from our model that would fall
within the masked aperture and then used this fraction to convert
between measured and total flux.  The aperture corrections used to
convert between measured and total flux are listed in Table
\ref{processedstuff}, column two.  The luminosities are presented in
units of $10^{43}$ erg\ s$^{-1}$ in Table \ref{processedstuff}, column
six.

\section{Sources of Error}

Clusters of galaxies represent only a small fraction of the
astronomical objects that emit X-rays. We have therefore investigated
how accidental coincidences between non-cluster sources and PDCS
cluster candidates effect the results in Table
\ref{processedstuff}. In Sec. 4.1, we calculate the number of
non-cluster objects expected to fall in our apertures using
established logN-logS relations. In Sec. 4.2, we highlight the
detections which are most likely to be of contaminating sources. In
Sec. 4.3, we compare our methodology to that of \cite{castander94}
(1994).  Finally, in Sec. 4.4, we discuss the surface density and
redshift distribution of X-ray emitting clusters in our sample as a
measure of our completeness.

\subsection{Expected contamination}

Using the logN-logS relation derived from the UK Deep and Medium
Surveys with ROSAT (\cite{brand94}), we were able to estimate the
number of non-cluster source detections in our survey.  For each of
the 31 PDCS cluster candidates, we used the 3$\sigma$ flux limit and
the integrated the logN-logS to compute the number of objects expected
per unit angular area.  To derive the area for each source, we used
the aperture size (Column 9 of Table \ref{rawdata} or 2\arcmin\,, if
this was smaller.  (Sources lying more than 2\arcmin\ from the PDCS
centroid are excluded from our apertures, see Sec. 3.)  We then
multiplied the area of aperture by the fraction of the area used in
the actual flux determination (Column 9 of Table \ref{rawdata}) to
determine angular area the aperture subtended.  The results from the
31 calculations were summed to give a contamination level of
$2.9^{+3.3}_{-1.4}$ sources (90\% confidence limits).  This level of
contamination is quite high and it is possible that all of the
detections flagged in Table \ref{processedstuff} correspond to
non-cluster sources.  We repeated the measurements of Section 3 but
with a cutoff radius of 1.5\arcmin\ or three times the estimated error
in the PDCS positions, instead of the above 2\arcmin.  Our calculated
error decreases to $2.1^{+2.4}_{-1.0}$ sources (90\% confidence
limits).  However, PDCS 01 is no longer considered a detection in this
sample.  This leaves the net number of detections approximately the
same.  Therefore, our estimate of the net number of X-ray detections
associated with PDCS cluster candidates appears to be insensitive to
the details of the X-ray analysis.

\subsection{Possible Identification of Non-cluster X-ray Sources}

We searched the NASA/IPAC Extragalactic Database (NED, \cite{ned}) for
possible sources of non-cluster X-ray emission near to our six
detections (PDCS 01, 33, 36, 61, 62, \& 63).  For PDCS 01, nothing was
found within 5\arcmin\ of the PDCS position.  A radio source is
located 2\arcmin\ away from PDCS 33, but this object is not coincident
with the detected X-ray emission.  For PDCS 36, the optical and X-ray
centroid are coincident and there are no nearby likely sources of
X-ray emission listed in NED.  There are three FIRST
(\cite{firstpaper}) radio sources within 5\arcmin\ of PDCS 61 and all
of these lie in the vicinity of the strong X-ray emission to the north
west of the PDCS centroid.  An elliptical galaxy lies 3.3\arcmin\ from
PDCS 62, but this object is further from the X-ray centroid than the
cluster candidate (the latter have a separation of only 0.9\arcmin).
For PDCS 63, no nearby optical or radio source was found in NED.
Based on this search, and on the comparison of optical and X-ray
positions, we find it probable that the cluster candidate is the
source of the detected X-ray emission in the case of PDCS 01, 36, 62,
\& 63. The other detections, around PDCS 33 \& 61, might be due to
contaminanting sources. However, these results are subjective and
cannot be confirmed without optical spectroscopy.  Therefore, for the
remainder of this paper, we will assume the contamination levels
derived statistically in Sec. 4.1.

\subsection{Comparison with Castander \etal 1994}

In this section, we compare the results of our X-ray analysis to those
from an X-ray survey by \cite{castander94} (1994) of high redshift GHO
clusters (\cite{gunn86} 1986). We have two clusters in common with
this survey, PDCS 59 and 63. For PDCS 63, or Cl1322+3029,
\cite{castander94} (1994) reported a $3.0\sigma$ detection and a
luminosity of $7.3 \pm 2.2 \times 10^{43}$ (0.1 - 2.4 keV) erg
s$^{-1}$.  We measure a $4.7\sigma$ detection and a
luminosity of $L_{x} = 14.0 \pm 3.0 \times 10^{43}$ (0.4 - 2.0 keV)
erg s$^{-1}$.  We attribute this difference to the bright X-ray source
which lies 2.5\arcmin\ from PDCS 63, see Figure 1. The manner in which
this source is masked from the detection and background apertures
strongly effects the flux measured for PDCS 63. The difference between
our luminosity measurement for PDCS 63 and that of \cite{castander94}
(1994) is not significant since the statistical error we quote most
probably underestimates the uncertainty in the measurements.
 
Our results for PDCS 59, or Cl1322+3027, also differ.
\cite{castander94} (1994) report a $3.1\sigma$ detection while we have
a $2.0\sigma$ result.  We have found that this discrepancy is largely
a result of the differences between the software package (EXAS) we
used compared to that (SASS) used by \cite{castander94} (1994).  Most
of the difference between our result and the result of
\cite{castander94} (1994) stems from the rejection of time intervals
by EXAS where the threshold count rate of 5.7 cts s$^{-1}$ was
exceeded (see Sec. 3 for details).  Using the SASS produced maps
instead of the EXAS software, we measure a $3.0\sigma$ flux for PDCS
59.

For PDCS 59, \cite{castander94} (1994) measure a luminosity which is
1$\sigma$ below the 3$\sigma$ upper limit quoted in Table
\ref{processedstuff}; $L_{x} = 6.7 \pm 2.1 \times 10^{43}$ (0.1 - 2.4
keV) erg s$^{-1}$ compared to $L_{x} = 9.4 \times 10^{43}$ (0.4 - 2.0
keV) erg s$^{-1}$. Our measured error is 1.5 times higher than that
measured by \cite{castander94} (1994) and we believe this is a
consequence of the EXAS removal of events falling in bad time
intervals. Whether or not PDCS 59 meets the criteria for detection
depends sensitively on how we calculate the background and how we
attempt to eliminate systematic errors.  For the remainder of the
paper we will make the conservative assumption that PDCS 59 was not
detected.

As our results differ with \cite{castander94} (1994) for both PDCS 59
and 63, we decided to apply our X-ray analysis to 3 other clusters, Cl
1322+3115, Cl 1603+4313, and Cl 1603+4329. These clusters are not in
the PDCS, but the relevant, EXAS reduced, PSPC pointings were
available to us via the SHARC Survey. We conducted the same analysis
discussed in Sec. 3 on these clusters, and in all three cases we find
our error to be higher than that published in \cite{castander94}
(1994). Despite this, we agree with \cite{castander94} (1994) as to
the status of these clusters; we measure only upper limits for Cl
1322+3115 and Cl 1603+4329 and, for Cl 1603+4313, we register a
$4.7\sigma$ detection. (\cite{castander94} 1994, quote a similar -
$4.3\sigma$ - detection for this cluster.) In conclusion, we have
found that our X-ray analysis does not differ significantly from that
of \cite{castander94} (1994) and that we agree on the the detection or
non-detection of 4 out of 5 objects in the \cite{castander94} (1994)
sample.  We do find a systematic difference in the measurement of the
error for our sources. This causes a corresponding increase in the
measured luminosity or luminosity upper limit.  However, most of the
apparent dissimilarities are at the $1-3\sigma$ level.  These are
attributable to the use of differing software packages and background
measurement techniques.
  
\subsection{Surface Density and Completeness}

The angular overlap between the 14 PSPC pointings we used and the 5
PDCS survey fields was 1.85 deg$^{2}$. The majority of this area, 1.38
deg$^{2}$, was surveyed to a flux limit of $\simeq 3 \times 10^{-14}$
ergs s$^{-1}$ cm$^{-2}$ (0.4 - 2.0 keV).  We have derived the surface
density of X-ray clusters in this 1.38 deg$^{2}$ area and compared it
to previous results. Of the six detections listed in Table
\ref{processedstuff}, five lie in this study area.  We have repeated
the calculation described in Sec. 4.1 for this area and flux limit.
We derive an expectation value of $2.64^{+3.02}_{-1.26}$ chance
coincidences of non-cluster X-ray sources with PDCS cluster
candidates.  The net number ($2.36^{+1.26}_{-3.02}$) of X-ray emitting
clusters over the 1.38 deg$^{2}$ area yields a surface density of
$1.71^{+0.91}_{-2.19}$ clusters deg$^{-2}$ to a limiting flux of
$\simeq 3.0\times 10^{-14}$ ergs s$^{-1}$ cm$^{-2}$ (0.4 - 2.0 keV).
A number of surveys have been conducted to similar limiting fluxes
over areas on the order of 15 deg$^{2}$ (\cite{castander95} 1995,
\cite{rosati95}, \cite{scharf97}, \cite{collins97} 1997).  Although
not all of these groups have completely identified their sources, most
report between 2 and 3 clusters deg$^{-2}$. The lowest reported
surface density is $\sim 1$ deg$^{-2}$ from \cite{castander95} (1995).
Our result, of $1.71^{+0.91}_{-2.19}$ clusters deg$^{-2}$, is
therefore in good agreement with other surveys, especially considering
the large error on our contamination rate (see Sec. 4.1).  We conclude
that it is unlikely that our survey has missed any X-ray emitting
clusters, though we cannot rule out that possibility.

Using the photometrically estimated redshifts from the PDCS, we find
the range of luminosities for the detected PDCS candidates to be $0.34
< L_x < 2.00 \times 10^{44}$ erg s$^{-1}$, with $\bar{L_x} = 0.9\times
10^{44}$ ergs s$^{-1}$ (0.4 - 2.0 keV).  These values are consistent
with those derived from other X-ray cluster surveys constructed from
PSPC pointings to similar flux limits.  Assuming the PDCS photometric
redshifts are not systematically biased, the redshift distribution of
the six candidate X-ray clusters in our sample roughly agrees with
that measured by \cite{collins97} (1997). The redshift range of these
6 objects is $0.3<z<0.69$, with $\bar{z}=0.46$.  We conclude that the
objects in our survey are similar, in terms of surface density,
luminosity range and redshift distribution, to those found in X-ray
selected distant cluster samples.

\section{Discussion}

Our X-ray survey of 31 PDCS cluster candidates has resulted in 6
detections and 25 upper limits. We have combined both the detections
and the upper limits to study the relation between X-ray luminosity
and optical richness for PDCS clusters. We have compared this relation
to that derived from lower redshift samples to understand what sort of
galaxy overdensities were selected by the PDCS.

In Figure \ref{lambdalx} we plot the richness, $\Lambda_{cl}$,
measured from the PDCS (see Sec. 2.)  versus the estimated X-ray
luminosity, $L_x$, corresponding to the detections and upper limits.
As the PDCS was conducted in two different passbands, each cluster
candidate has two values for $\Lambda_{cl}$ and both are shown in the
figure.  Using the techniques outlined in \cite{isobe86} (1986) as
implemented in the STSDAS survival analysis package, we can test our
data for a correlation.  We measure a correlation between $L_x$ and
richness at the 96.6\% confidence limit for the $V_4$ band and at the
93.2\% confidence limit for $I_4$ band. This is in contrast to the
strong correlation, at the $\geq 99.95\%$ confidence level, measured
for Abell clusters (\cite{abram83}, \cite{kowalski84} 1984;
\cite{bh93} 1993). We note that the analysis of \cite{kowalski84}
(1994), like that used here, combined both detections and upper
limits. The lack of a statistically significant correlation
($>99.95\%$) in our data could be a reflection of several systematic
biases in our sample; contamination by non-cluster X-ray sources, the
error in the PDCS distance estimates, or errors in our flux
measurements.

We have also plotted on Figure \ref{lambdalx} (solid line), the fitted
$L_x$ versus richness relation for Abell clusters taken from
\cite{bh93} (1993). We have adjusted this relation to take account of
the differing passbands of our work compared to that of the ROSAT All
Sky Survey, but have assumed that the PDCS richness ($\Lambda_{cl}$)
is equal to the Abell richness ($N_{Abell}$, see Sec. 2).  The dashed
lines on Figure \ref{lambdalx} represent the median luminosities for
Richness class 0, 1, \& 2 Abell clusters taken from \cite{burg94}
(1994), after appropriate conversion from the Einstein IPC detector
passband (0.5 - 4.5 keV).  Although care must be taken in interpreting
Figure \ref{lambdalx}, since the majority of our data points are upper
limits, it does seem to illustrate that our data have a similar X-ray
luminosity distribution to that of low redshift clusters.

One of the largest sources of uncertainty in our luminosity versus
richness analysis is the error in the estimated redshift, which can be
as large as $\delta_z \sim 0.2$. Therefore, we now consider a redshift
independent quantity that measures the correlation between optical and
X-ray luminosities.  We used the ratio of the optical luminosity to
the X-ray luminosity ($L_{opt}/L_{x}$), where $L_{opt}$ is defined as
the optical richness, $\Lambda_{cl}$, multiplied by the absolute
luminosity of an $L_\star$ galaxy in the $V_4$ band, as specified in
Table 3 of \cite{postman96} (1996). The resulting ratio is not truly
redshift independent, since the apertures used to construct the
optical richness and X-ray luminosity values require a redshift
estimate, but redshift dependence has been minimized.  This method has
been used previously by \cite{stocke91} (1991) to measure
$L_{opt}/L_{x}$ ratios for EMSS clusters, though we note that the
ratio we use here is slightly different. Whereas we define optical
luminosity using the bright end of the cluster member luminosity
function, \cite{stocke91} (1991) relied only on the magnitude of the
brightest cluster galaxy.

We plot, in Figure \ref{lamratio}, the $L_{opt}/L_{x}$ ratio as a
function of estimated redshift for the 31 PDCS cluster candidates in
our sample.  We also plot the $L_{opt}/L_{x}$ ratio for the Abell
clusters in the \cite{bh93} (1993) survey. For these clusters,
$L_{opt}$ is given by the Abell richness, $N_{Abell}$, multiplied by
$L_\star$.  (Both $\Lambda_{cl}$ and $N_{Abell}$ correspond to the
number of $L_\star$ galaxies in a 3.0 Mpc radius.)  From Figure
\ref{lamratio}, it is apparent that the $L_{opt}/L_{x}$ ratios of the
detections in our survey are similar to those of Abell clusters in
\cite{bh93} (1993).  We have performed a number of statistical tests
(the Gehan, logrank, Peto-Peto, and the Peto-Prentice tests, see Sec.
III of \cite{schmitt85} 1985 for a discussion of statistical tests
comparing two samples with censoring) using the STSDAS survival
analysis package to compare the distributions of the ratio for the 2
samples.  Although it might appear that the ratio of X-ray luminosity to
optical luminosity is, on average, slightly lower in the PDCS sample,
we found no statistically significant ($>$ 95\%) difference between
the PDCS sample and the sample of \cite{bh93} (1993) in any of the
four different tests we tried. We note that the apparent downturn in
the $L_{opt}/L_{x}$ ratio for the PDCS sample at $z\gtrsim 1.0$,
coincides with the redshift at which the PDCS catalog becomes
incomplete.

In summary, for the 31 PDCS cluster candidates in our sample, we do
not measure a statistically significant correlation between $L_x$ and
richness. Comparisons with the \cite{bh93} (1993) and \cite{burg94}
(1994) surveys, illustrate that our clusters are not dissimilar - in
terms of their luminosity range at a given richness - to Abell
clusters at lower redshifts. Using a redshift independent test
involving optical to X-ray luminosity ratios, we have found no
significant difference between PDCS and Abell clusters.  Both samples,
PDCS and Abell, demonstrate a wide range in optical to X-ray
luminosities at a given redshift or optical richness.  

\section{Conclusions}

We have searched for X-ray emission around 31 distant, optically
selected, rich clusters of galaxies taken from the PDCS sample of
\cite{postman96} (1996). We have found six possible coincidences
between PDCS clusters and X-ray sources detected in ROSAT/PSPC
pointings. Our statistical analysis, and explicit searches for
alternative optical or radio counterparts to the X-ray sources, are
consistent with 3 of these 6 being detections of contaminating,
non-cluster, sources.
 
Our survey demonstrates that PDCS cluster candidates are not strong
($L_x\gtrsim 5 \times 10^{43}$ erg s$^{-1}$) X-ray emitters.  This
observation is consistent with the results of previous studies of
distant, optically selected, clusters (\cite{bower94} 1994;
\cite{castander94} 1994; \cite{nichol94} 1994; \cite{sok96}) and
highlights the possible existence of a population of optically rich
clusters which are X-ray faint. The average luminosity of the six
detections in our survey, $\bar{L_x} = 0.9\times 10^{44}$ ergs
s$^{-1}$, and the estimated surface density of X-ray emitting clusters
in the PDCS fields, 1.71 deg$^{-2}$, are consistent with the results
of previous surveys for distant, X-ray selected, clusters
(\cite{castander95} 1995; \cite{rosati95}; \cite{scharf97};
\cite{collins97} 1997).

Taking both the upper limits and the detections, we find no
statistically significant correlation between optical richness and
X-ray luminosity in our sample, in contrast to the strong correlation
($\geq 99.95\%$ confidence level) measured previously for Abell
clusters (\cite{abram83}; \cite{kowalski84} 1984; \cite{bh93} 1993).
This is not surprising given the large uncertainties in the redshifts
of the PDCS.  We find that our sample is in agreement with that of
\cite{bh93} (1993) when we compare optical to X-ray luminosity ratios,
which is a redshift independent test.  Our data are, therefore,
consistent with the hypothesis that the PDCS and the Abell catalog
sample the same section of cluster population, albeit at different
redshifts. Furthermore, we find no evidence for evolution in the X-ray
versus optical properties of rich clusters out to a redshift of
$z=0.6$.

The combination of the X-ray data presented here with the objective
selection of the PDCS, makes this sample of 31 distant clusters ideal
for further studies of cluster evolution. We are actively engaged in
following up this sample with photometry and spectroscopy. These new
data will allow us not only to determine which cluster candidates
represent real physical systems, but also to estimate the volume
density of X-ray emitting clusters and to probe the relationship
between velocity dispersion and X-ray luminosity.  These various
studies will allow us to quantify the degree to which cluster
properties are evolving with redshift.

\section{Acknowledgments}

We are extremely grateful to many people for stimulating discussions
during this work. We thank the Lori Lubin, Constance Rockosi,
Francisco Castander and Marc Postman for their time and helpful
comments.  We would like to especially thank Rich Kron for his careful
readings of earlier manuscripts and suggested improvements in the
analysis.  Finally, we thank an anonymous referee for their careful
reading of the paper and their constructive comments.  BH was
supported in part by the National Science Foundation under a
cooperative agreement with the Center for Astrophysical Research in
Antarctica (CARA), grant number NSF OPP 89-20223. CARA is a National
Science Foundation Science and Technology Center. AR and MU
acknowledge support from NASA ADP grant NAG5-2432. This research has
made use of data obtained through the High Energy Astrophysics Science
Archive Research Center Online Service, provided by the NASA-Goddard
Space Flight Center.  This research has made use of the NASA/IPAC
Extragalactic Database (NED) which is operated by the Jet Propulsion
Laboratory, Caltech, under contract with the National Aeronautics and
Space Administration.

\newpage

\epsscale{0.7}
\plotone{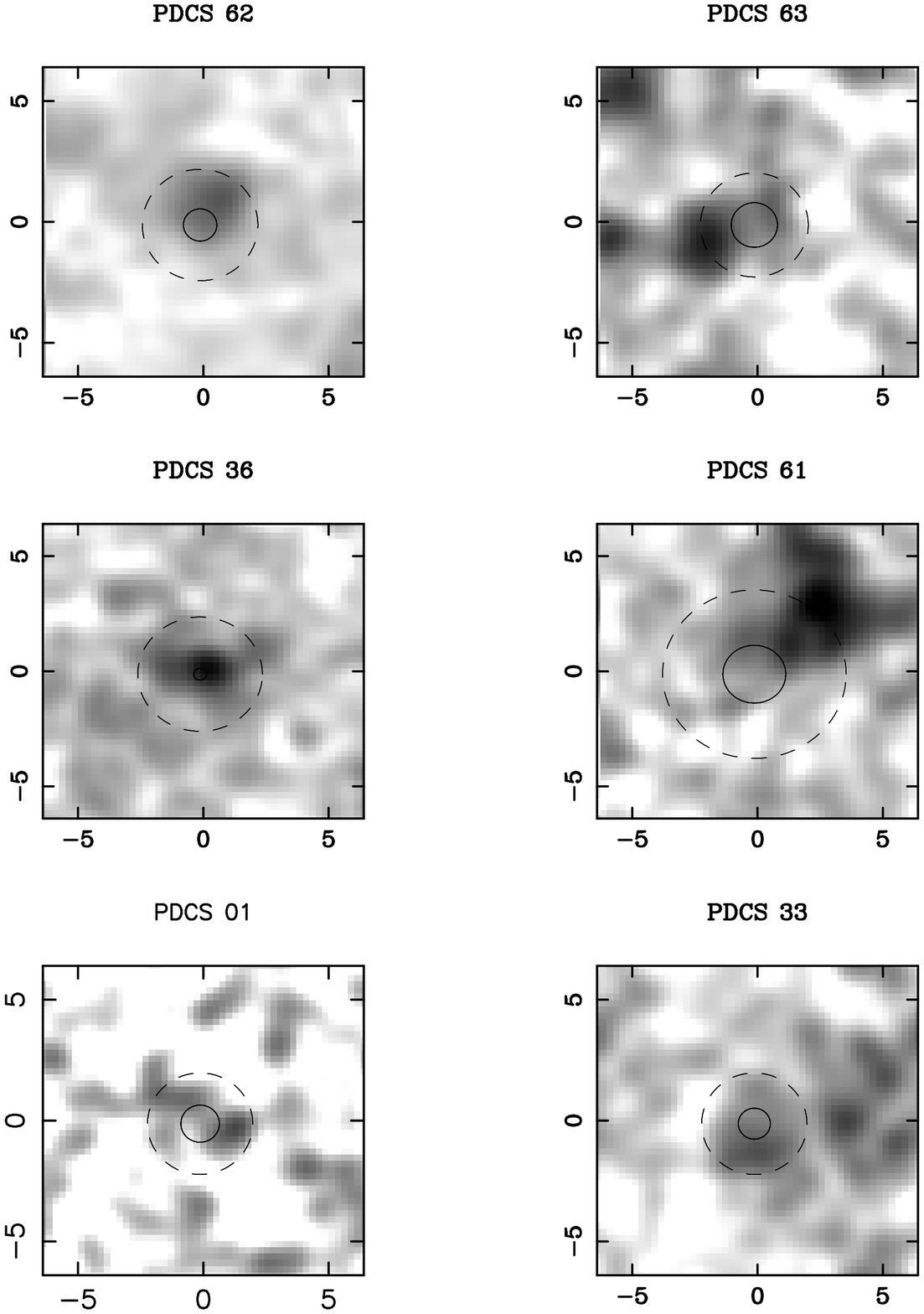}

\figcaption[holden.fig1.ps]{X-ray images of the six
PDCS cluster candidates in our survey that are coincident with
$>3\sigma$ detections. Each image is 12.7\arcmin\ on a side and
centered on the PDCS optical centroid.  The darkest grey scale
represents \(1.7 \times 10^{-14} \) ergs s$^{-1}$ cm$^{-2}$
arcmin$^{-2}$, while the lightest grey scale represents \(2.7 \times
10^{-15} \) ergs s$^{-1}$ cm$^{-2}$ arcmin$^{-2}$, with logarithmic
scaling between these values.  For PDCS 62, the darkest greyscale
represents \(1.07 \times 10^{-13} \) ergs s$^{-1}$ cm$^{-2}$
arcmin$^{-2}$.  The inner circle, represented by a solid line, is the
half light radius of the corresponding off-axis point spread
function. The outer circle, a dashed line, shows the unmasked aperture
used to calculate the flux for the object.  The tick marks represent 5
arcminutes of angle.
\label{mosaic}}

\epsscale{0.7}
\plotone{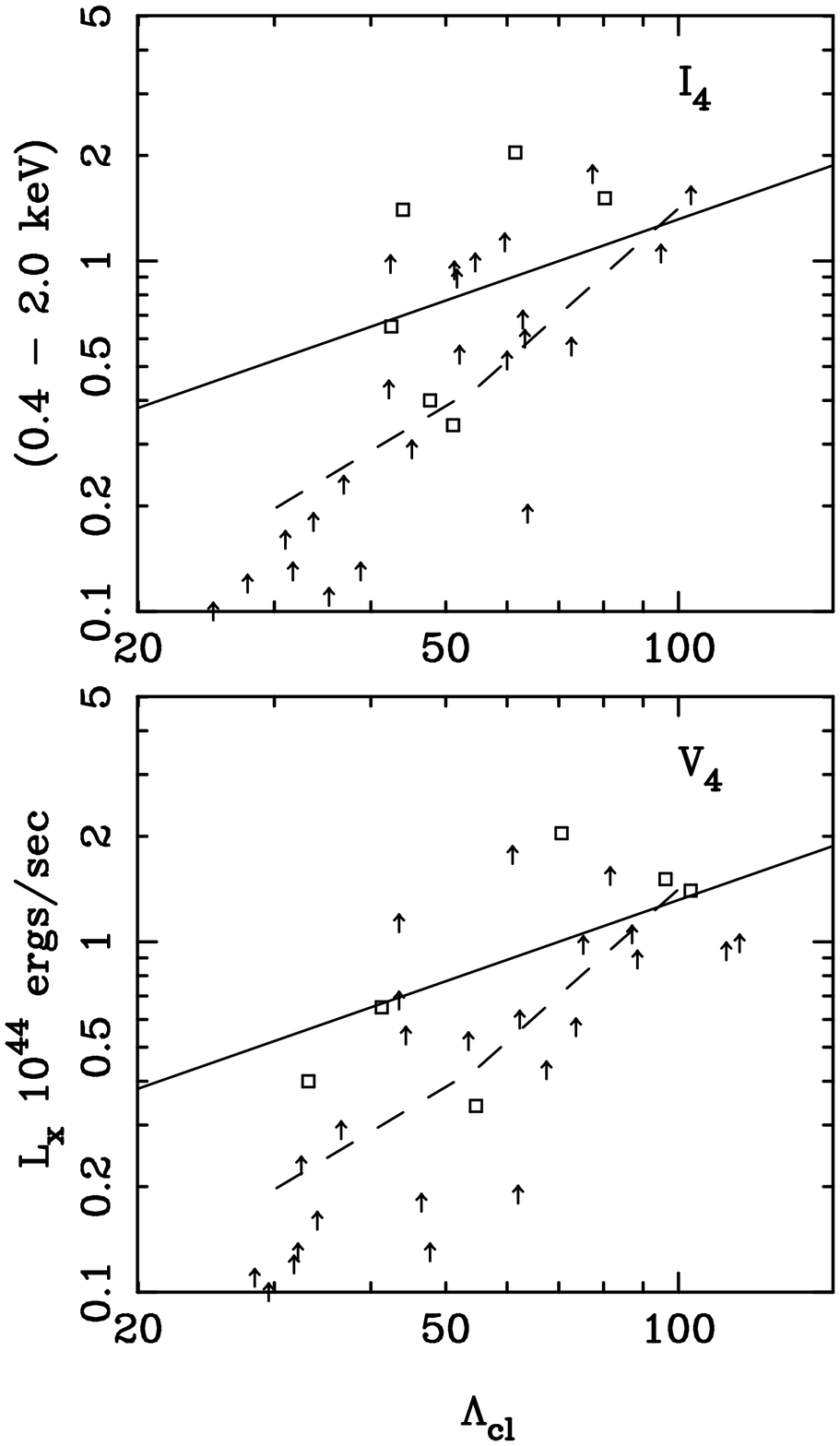}

\figcaption[holden.fig2.ps]{The $\Lambda_{cl}$ Richness versus X-ray
luminosity for the 31 PDCS cluster candidates in our sample. The
squares are the detections and the arrows represent the $3\sigma$
upper limits. The solid line represents the fit to the $L_x$ versus
richness relation for Abell clusters taken from Briel \& Henry
(1993). The dashed line segments  represent the
median luminosity for Richness class 0, 1 \& 2 Abell clusters
presented in Burg \etal (1994). The lines have been adjusted to the
passband of our survey and assume that $\Lambda_{cl}$ is equivalent to
the Abell richness.  The top plot uses the richness values from the
$I_{4}$ band while the bottom plot uses the richness values from the
$V_{4}$ band.
\label{lambdalx} }

\epsscale{0.7}
\plotone{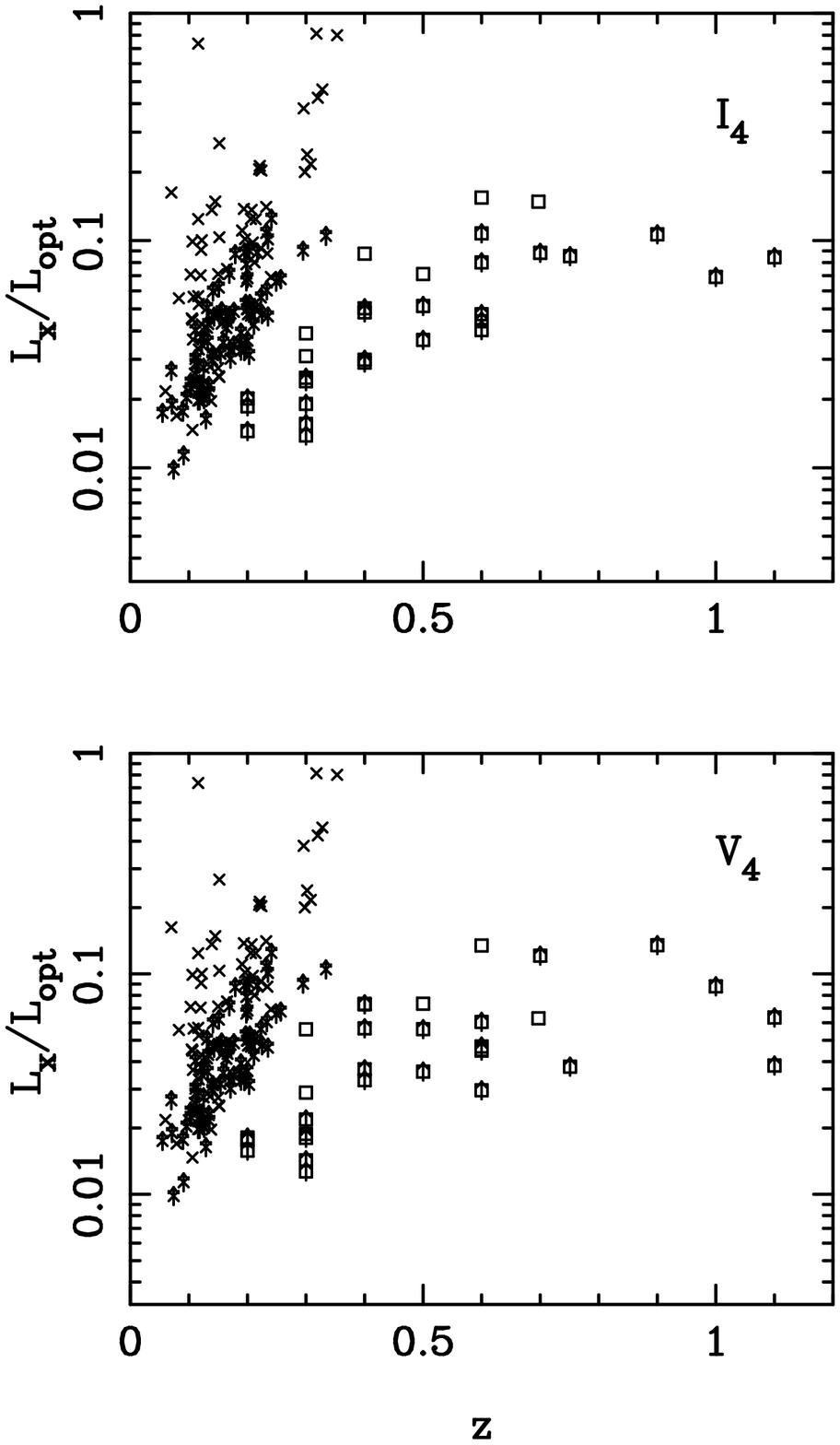}

\figcaption[holden.fig3.ps]{The optical luminosity to X-ray luminosity
($L_{opt}/L_x$) ratio versus redshift for the 31 PDCS cluster
candidates in our survey and for the Abell clusters in the sample of
Briel \& Henry (1993).  The squares are from our sample and the
crosses are from the sample of Briel \& and Henry 1993.  Arrows
through a symbol represent the $3\sigma$ upper limits from both
samples.  The top plot uses the richness values from the $I_{4}$ band
while the bottom plot uses the richness values from the $V_{4}$
band. \label{lamratio} }

\begin{table}
\dummytable\label{pointings}
\end{table}
\begin{table}
\dummytable\label{rawdata}
\end{table}
\begin{table}
\dummytable\label{processedstuff}
\end{table}

\end{document}